# IoT based Personal Voice Assistant

Research Paper

[1]Sumit Kumar, [2]Sankalp Sagar, [3]Varun Gupta, [4]Sachin Kumar Singh

[1,2,3,4]Department of CSE, Maharana Pratap Engineering College, Kanpur

*Abstract* - **Today the technological advancement is increasing day by day. Earlier only there was a computer system in which we can do only few tasks. But now machine learning, artificial intelligence, deep learning, and few more technologies have made computer systems so advance that we can perform any type of task. In such era of advancement if people are still struggling to interact using various input devices, then it's not worth it. For this reason, we developed a voice assistant using python which allows the user to run any type of command in Linux without interaction with keyboard. The main task of voice assistant is to minimize the use of input devices like keyboard, mouse etc. It will also reduce the hardware space and cost.**

## INTRODUCTION

In this era of technology everything that human being can do are being replaced by machines. One of the main reasons is change in performance. In today's world we train our machine to think like humans and do their task by themselves. Therefore, there came a concept of virtual assistant.

A virtual assistant is a digital assistant that uses voice recognition features and language processing algorithms to recognize voice commands of user and perform relevant tasks as requested by the user. Based on specific commands given by the user a virtual assistant is capable of filtering out the ambient noise and return relevant information.

Virtual Assistant are completely software based but nowadays they are integrated in different devices and also some of the assistants are designed specifically for single devices like Alexa.

Due to drastic change in technology now it's a. high time to train our machine with the help of machine learning, deep learning, neural networks. Today we can talk to our machine with the help of Voice Assistant. Today every big company is using Voice Assistant so that their user can take the help of machine through their voice. So, with the Voice Assistant we are moving to the next level advancement where we are able to talk to our machine. These types of virtual assistants are very useful for old age, blind & physically challenged people, children, etc. by making sure that the interaction with the machine is not a challenge anymore for people. Even blind people who couldn't see the machine can interact with it using their voice only.

Here are some of the basic tasks that can be done with the help of voice assistant: -
- Reading Newspaper

- Getting updates of mail
- Search on web
- Play a music or video.
- Setting a reminder and alarm
- Run any program or application.
- Getting weather updates

These are some of the examples, we can do many more things according to our requirement. The Voice Assistant that we have developed is for Windows users as well as for Linux Users. The voice assistant we have developed is a desktop based built using python modules and libraries. This assistant is just a basic version that could perform all the basic tasks which have been mentioned above but current technology is although good in it is still to be merged with Machine Learning and Internet of Things (IoT) for better enhancements.

We have used python modules and libraries for making the model and we have used Machine Learning for training our model, some of the windows and Linux commands are also added to model so that our model can run smoothly on this operating system.

Basically, our model will work in three modes:
1. Supervised Learning
2. Unsupervised Learning
3. Reinforcement Learning

Depending upon the usage for which the assistant is required for user. And these can be achieved with the help of Machine learning and Deep Learning. With the help of Voice Assistant there will be no need to write the commands again and again for performing particular task. Once model is created it can be used any number of times by any number of users in the easiest ways.

So, with the help of virtual assistant, we will be able to control many things around us single handedly on one platform.

## II. LITERATURE SURVEY

Bassam A, Raja N. et al, written about statement and speech which is most significant. In the communication between human and machine arrangement was done through analog signal which is converted by speech signal to digital wave. This technology is massively utilized, it has limitless uses and permit machines to reply appropriately and consistently to user voices, also offers useful and appreciated facilities. Speech Recognition System (SRS) is rising gradually and has indefinite applications. The research has revealed the summary of the procedure; it is a simple model [1].

B. S. Atal and L. R. Rainer et al, explained regarding speech analysis, and result is regularly completed in combination with pitch analysis. The research described a pattern recognition technique for determining whether a given slice of a speech signal should be categorized as voiced speech, unvoiced speech, or silence, depending on dimensions finished on signal. The main restriction of the technique is the requirement for exercise the algorithm on exact set of dimensions picked, and for the specific recording circumstances [2].

V. Radha and C. Vimala et al, explained that most general mode of communication among human beings is speech. As this is the utmost technique, human beings would identical to utilize speech to interrelate with machines too. Because of this, autonomous speech identification has got a lot of reputation. Most techniques for speech recognition be like Dynamic Time Warping (DTW), HMM. For the feature mining of speech Mel Frequency Cestrum Coefficients (MFCC) has been utilized which offers a group of characteristic vectors of speech waveform. Prior study has exposed MFCC to be more precise and real than rest characteristic mining approaches in the speech recognition. The effort has been completed on MATLAB and investigational outcomes depict that system is capable of identifying words at satisfactorily great accuracy [3].

T. Schultz and A. Wail et al, explained about the spreading of speech technology products around the globe, the immovability to novel destination languages turns out to be a useful concern. As a significance, the research emphases on the query of how to port huge vocabulary incessant speech recognition (LVCSR) systems in a fast and well-organized manner. More particularly the research needs to evaluate acoustic models for a novel destination language by means of speech information from different source languages, but only restricted data from the destination language identification outcomes using language-dependent, independent and language-adaptive acoustic models are described and deliberated in the framework of Global Phone project which examines LVCSR methods in 15 languages.[4].

J. B. Allen et al described about the Language that is the utmost significant means of communication and speech is its major interface. The interface for human to machine, speech signal was converted into analog and digital wave shape as a machine understood. [10] A technology enormously utilized and has limitless applications. Speech technologies permit machines to react appropriately and consistently to human speeches and offers valuable and appreciated services. The research provides a summary of the speech identification procedure, its basic model, and its application, techniques and also describes reasonable research of several techniques that are utilized for speech recognition system. SRS is enhancing gradually and has infinite applications. [5]

Mughal Bapat, Pushpa Bhattacharyya et al, described a morphological analyzer for most of the NLP solicitations of Indian Languages. [11] During the work they described and estimated the morphological analyzer for Marathi language. They started by planning a to some extent homomorphism "boos trappable" encryption technique that functions during the function f is the techniques individual decryption function. The research showed a great accuracy for Marathi that adventures consistency in inflectional standards in engaging the Finite State Systems for demonstrating language in a sophisticated way. Grouping of post positions and the growth of FSA is one of significant assistances since Marathi have difficult morphotactic [6].

G. Muhammad, M. N. Huda et al, presented a model ASR for Bangla digit. Although Bangla is among the mostly spoken languages around the globe, some of the few works of Bangla ASR can be identified in the collected works, particularly Bangla accented in Bangladeshi. During this research, the quantity is gathered from publics in Bangladesh. Mel-frequency cepstral coefficients (MFCCs) dependent characteristics and hidden Markov model (HMM) dependent classifiers are utilized for identification. Dialectical variance make happen a part of performance deprivation. In situation of gender-based trials, female spoken digits had greater accuracy rates than those by male spoken digits [7].

Sean R Eddy et al operated on Hidden Markov models which are a common statistical designing approach for 'linear' issues like sequences or time series and have been extensively utilized in speech identification requests for twenty years. Inside the HMM formalism, it is probable to relate formal, completely probabilistic techniques to profiles and gapped structure arrangements.[12] Profiles based on Hidden Markov model have fixed most of the concerns related with typical profile analysis. HMMs offer a steady theory for notching insertions and deletions, and a constant structure for joining structural and sequence data. HMM based numerous sequence arrangements is quickly refining. Homolog recognition based on HMM is previously adequately influential for HMM techniques to relate satisfactorily to much more difficult threading techniques for protein reverse fold [8].

### III. PROBLEM FORMULATION

This section describes the description about the problem formulation.

As we know each human have their own characteristics and every developer applies his own method and approaches for development of a product. One assistant can synthesize speech more qualitatively, another can more accurately and without additional explanations and corrections perform tasks, others are able to perform a narrower range of tasks, but most accurately and as the user wants. Therefore, there is no such assistant that can perform all the work and tasks equally. The set of characteristics that an assistant has depends on the area on which developer paid more attention. Since all system are based on machine learning and use for their creation huge amounts of data collected from various sources and then trained on them, an important role is played by the source of this data.

Despite the different approaches to learn different algorithms, the principle of building voice assistant remains the same. The technologies that are used to build a voice assistant that can interact with the humans are speech recognition, Teach-To-Speech, voice biometrics, dialog manager, natural language understanding and named entity recognition.

| VOICE TECHNOLOGY | BRAIN TECHNOLOGY |
|---|---|
| Voice Activation | Voice Biometrics |
| Automatic Speech Recognition (ASR) | Dialog Management |
| Teach-To-Speech (TTS) | Natural Language Understanding (NLU) |
| | Named Entity Recognition (NER) |

### IV. PROPOSED APPROACH

The proposed system will have the following functionality:
(a) The system will keep listening for commands and the time for listening is variable which can be changed according to user requirements. (b) If the system is not able to gather information from the user input it will keep asking again to repeat till the desired no. of times.
(c) The system can have both male and female voices according to user requirements.
(d) Features supported in the current version include playing music, emails, texts, search on Wikipedia, or opening system installed applications, opening anything on the web browser, etc. (e) The system will keep listening for commands and the time for listening is variable which can be changed according to user requirements.
(f) If the system is not able to gather information from the user input it will keep asking again to repeat till the desired no. of times.
(g) The system can have both male and female voices according to user requirements [9].

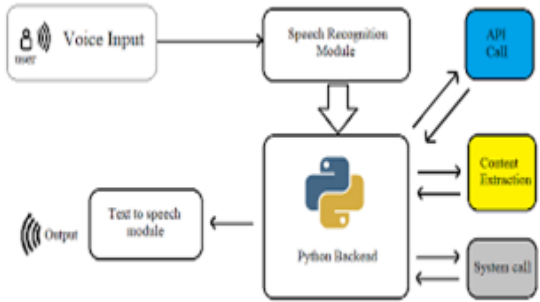

This is the Linux Code which will run on the server side for running the Linux command and displaying the output on the web.

## V. RESULT AND ANALYSIS

This section describes a brief description of our result which will be displayed is to start speaking. After this on the basis of the comparison and analysis of our proposed work. We have employed this idea by means of Python, Machine Learning and AI. Our main aim is to assist the users in their tasks with the help of their voice commands. This can be done in two phases. Firstly, taking the audio input from the user and converting it to an English phrase with the help of Speech Recognition API. Secondly searching for the task user wants to perform and then redirecting it to the Linux server with the help of HTTP Protocol and displaying the result on the web browser.

When the Windows Code is executed the first Output is to start speaking. After this the user has to give the voice command.

This screen will be visible when user has given voice command and the Google Speech Recognition API has translated it into an English Phrase.

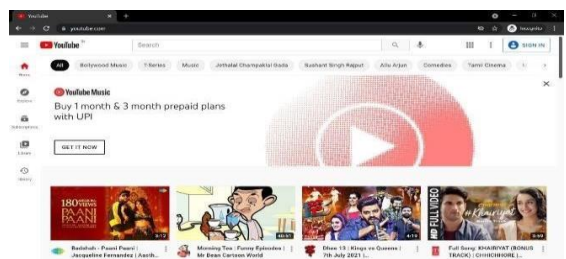

This is the Windows Code which will run on the client side for taking voice input of the user. After translation the command which the user has given will be displayed on the web browser.

## VI. CONCLUSION

In this paper we have discussed a Voice Assistant developed using python. This assistant currently works as an application based and performs basic tasks like weather updates, stream music, search Wikipedia, open desktop applications, etc. The functionality of the current system is limited to working on application based only. The upcoming updates of this assistant will have machine learning incorporated in the system which will result in better suggestions with IoT to control the nearby devices similar to what Amazon's Alexa does.